\documentclass[10pt,letterpaper,twocolumn]{article} 

\usepackage[usenames,dvipsnames]{color}
\usepackage{ol2}
\usepackage{amsmath,array,amssymb}
\usepackage{graphicx}
\usepackage{hyperref}
\usepackage{textcomp}
\usepackage{braket}
\usepackage{float}
\usepackage{multirow}
\usepackage{indentfirst}
\usepackage[FIGTOPCAP,raggedright,nooneline,bf,footnotesize]{subfigure}
\usepackage{paralist}
\usepackage{overpic}
\usepackage{gensymb}

\renewcommand{\eqref}[1]{Eq.~(\ref{#1})}

\twocolumn

\begin{document}
\twocolumn[ 
\title{Effects of fabrication methods on spin relaxation and crystallite quality in Tm-doped Y$_2$Al$_5$O$_{12}$ powders studied using spectral hole burning
}
\author{Thomas Lutz$^{1}$, Lucile Veissier$^{1}$, Charles W. Thiel$^{2}$, Philip J. T. Woodburn $^{2}$, Rufus L. Cone$^{2}$, Paul E. Barclay$^{1}$ and Wolfgang Tittel$^{1}$}
\address{
$^1$Institute for Quantum Science and Technology, and Department of Physics \& Astronomy, University of Calgary, Calgary Alberta T2N 1N4, Canada
\\
$^2$Department of Physics, Montana State University, Bozeman, MT 59717 USA\\
$^*$Corresponding author: thomasl@ucalgary.ca
}

\begin{abstract}
High-quality rare-earth-ion (REI) doped materials are a prerequisite for many applications such as quantum memories, ultra-high-resolution optical spectrum analyzers and  information processing. Compared to bulk materials, REI doped powders offer low-cost fabrication and a greater range of accessible material systems. Here we show that crystal properties, such as nuclear spin lifetime, are strongly affected by mechanical treatment, and that spectral hole burning can serve as a sensitive method to characterize the quality of REI doped powders.  We focus on the specific case of thulium doped Y$_2$Al$_5$O$_{12}$ (Tm:YAG). Different methods for obtaining the powders are compared and the influence of annealing on the spectroscopic quality of powders is investigated on a few examples. We conclude that annealing can reverse some detrimental effects of powder fabrication and, in certain cases, the properties of the bulk material can be reached. Our results may be applicable to other impurities and other crystals, including color centers in nano-structured diamond.
\end{abstract}

] 

\maketitle
\newpage

\section{Introduction}

\noindent Rare-earth-ion (REI) doped bulk crystals cooled to cryogenic temperatures are used for a multitude of applications. Examples are quantum memories \cite{Riedmatten2015}, quantum information processing \cite{Saglamyurek2014}, and ultra-high-resolution optical spectrum analyzers \cite{Menager2001}. 
In contrast to REI doped bulk materials, powders offer low cost and rapid prototyping. Furthermore, the understanding of powders constitutes a first step towards nanofabrication of devices from these materials \cite{mcauslan_strong-coupling_2009,obrien_interfacing_2014,walther_high_2015,lutz_modification_2015}. However, despite much effort \cite{Hong1998,malyukin_single-ion_2003,Macfarlane2001,perrot_narrow_2013}, producing monodisperse powders with properties comparable to those of bulk materials remains challenging. 

Fabrication or manipulation of REI doped powders can induce stress, especially during grinding or milling, that creates strain in the crystal lattice \cite{Heitjans2007,Scholz2002}. In addition, impurities can contaminate the host matrix during synthesis. Strain and impurities often significantly impact the performance of the powders in both signal processing and more general luminescence applications. The goal of this work is to study REI doped powders at temperatures near 1.6 K and improve their properties to reach those of bulk materials. Towards that goal, we compared properties of powders that were synthesized chemically or milled down from larger crystals using either high-energy planetary ball mills or low-energy tumbling mills. We used scanning electron microscope (SEM) to determine the shape and size of the particles, x-ray diffraction (XRD) to analyze the composition and phase of the crystalline structure, and we employed sensitive spectral hole burning  (SHB) methods to probe variations in the optical decoherence dynamics, deduced from the spectral hole width, and the $^{169}$Tm nuclear spin relaxation dynamics, deduced from the spectral hole lifetime. We find that induced damage and strain in the crystal lattice, which does not affect XRD or SEM measurements, can produce large variations in the measured low-temperature dynamics of the powders that are observed using SHB techniques. Thus, SHB can serve as a quantitative characterization tool, complementing traditional techniques such as XRD, SEM or Raman scattering. Our results also demonstrate that mechanical and thermal treatment of REI doped crystals influences properties, such as the lifetime of nuclear spins, in a surprisingly strong way.  Consequently, SHB is a well-suited technique to reveal the presence of residual damage in powders due to fabrication and to evaluate the effectiveness of methods used to reduce strain and improve material quality, such as thermal annealing.


\section{Experiment}

\subsection{Tm:YAG}
\begin{figure}[]
\centering
\includegraphics[width=\columnwidth]{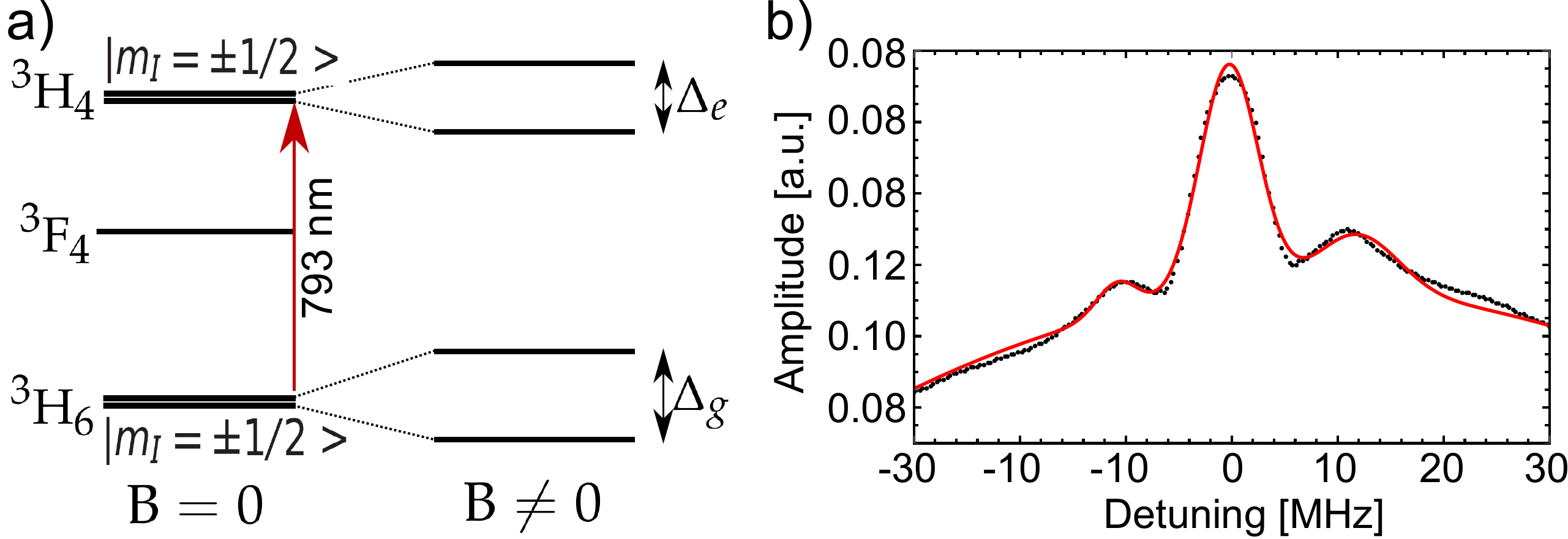}
\caption{(a) Level structure of Tm:YAG without and with an applied magnetic field. (b) Hole burning spectrum of the 1\% Tm:YAG bulk crystal together with a fit (red line).}
\label{fig:tm-yag-level-bulk}
\end{figure}

\noindent Our investigations employ crystalline Y$_3$Al$_5$O$_{12}$ doped with thulium impurities (Tm:YAG), whose relevant electronic level structure is shown in Figure \ref{fig:tm-yag-level-bulk}. Under an external magnetic field, both the $^3H_6$ ground state and the $^3H_4$ excited state split into two non-degenerate states through the hyperfine interaction with the $^{169}$Tm nuclear spin ($I=$ 1/2), allowing persistent atomic population storage with lifetimes as long as hours in bulk crystals at liquid helium temperatures through optical pumping of the nuclear spin states. 

The 20 GHz wide inhomogeneously broadened line of the $^3H_6$ $\leftrightarrow$ $^3H_4$ transition in Tm:YAG is centered at 793.156 nm \cite{sun_symmetry_2000}. The samples were mounted in an Oxford Instruments Spectromag cryostat and all powders were held in unsealed glass cuvettes. 
All samples were 0.5~mm thick, and, for the nominal Tm concentration of 1\%, featured an optical depth of $\approx$1. This allowed for direct transmission detection of spectral holes while keeping optical scattering at an acceptable level. To prevent scattered light from reaching the detector, the cuvette was placed inside a copper box with two pinholes to allow light to enter and exit. For all measurements, the samples were cooled in helium vapor to 1.6 K and the applied magnetic field was set to 1 T.


\subsection{Spectral hole burning}
\noindent To assess the quality of different powders, we used persistent spectral holes. We extracted the lifetime of the spin states from the hole decay, and the magnitude of optical decoherence from the width of the hole \cite{macfarlane_coherent_1987}. As we describe later, we found that both quantities strongly depend on the crystal quality in terms of strain as well as impurities. In this approach, a spectral hole is burnt into the inhomogeneously broadened absorption line by using a laser to optically pump population between the nuclear spin states. After a certain waiting time, the absorption line is scanned using a tunable laser and the area of the spectral hole is measured. From the exponential decay of the hole area with waiting time, the spin lifetime is extracted. Note that the width of the spectral hole ideally corresponds to twice the homogeneous linewidth; however, the measured width is generally larger due to laser frequency jitter, spectral diffusion, and power broadening. 

We used a Coherent 899-21 Ti:sapphire laser emitting at a wavelength of 793.38 nm and with a linewidth of less than 1 MHz. For the bulk single crystal measurement that served as a reference for the best material properties, the laser power was set to 5 $\mu$W. For the powder samples, more power was necessary to overcome loss due to scattering and achieve a sufficiently high signal-to-noise ratio. Otherwise, the experimental conditions were kept the same for all materials. The burn and read pulses were generated from the CW laser beam using two acousto-optic modulators (AOMs) in series. This arrangement gave an extinction ratio of $>$90 dB, ensuring that no unintentionally leaked light reached the sample. During the reading pulse, the laser sweep was implemented using a double-passed AOM scanned in frequency. A New Focus model 2051 photo-receiver was used to detect the transmitted light.

\subsection{Bulk single crystal reference}
\noindent First, as a reference against which we compare the properties of our Tm:YAG powders (the different methods used to create these powders are described below), we assessed the properties of a 1\% doped Tm:YAG crystal from Scientific Materials Corporation (SMC) (growth number 3-8). The crystal is 0.5 cm thick and features an optical depth of about 0.5. Depending on the crystal orientation, we found spin lifetimes between 5 and 16 hours  \cite{bulk-paper}. Furthermore, as shown in Figure \ref{fig:tm-yag-level-bulk} (b), the observed spectral hole width was limited to 6 MHz -- much wider than the intrinsic kHz-wide homogeneous linewidth of Tm:YAG \cite{Liu2005} under these conditions --primarily due to power broadening effects. The figure also shows clear side holes, split by 10 MHz/T and caused by super-hyperfine coupling between the thulium ions and the nuclear spin of the $^{27}$Al present in the host matrix \cite{macfarlane_coherent_1987,Bonarota2010}. Because powders are composed of randomly oriented crystallites, measurements of powders probe all possible orientations at once and we expect to observe lifetimes that span the range of those observed in the bulk material.

\section{Results}


\noindent In the following, we study the impact of fabrication and annealing methods on nuclear spin lifetimes T$_a$ as well as on hole linewidths $\Gamma$. Due to the strong scattering caused by the powders, higher laser power was required for these measurements, leading to additional power broadening. As a consequence, hole linewidths were measurable only down to approximately  10 MHz. A summary of the properties of all investigated materials is shown in Table \ref{tab:all}.

\begin{table}
\begin{tabular}{|l|c|c|c|}
\hline
	\bf{Material} & \bf{size} & \bf{$\Gamma$}& \bf{T$_a$}  \\ 
	& [$\mu$m]& [MHz]&[min]\\
	\hline\hline
	SMC Bulk 				& 			& \textbf{6}& 300-960\\ \hline 
	SMC thermal crushing: 	&500 	& $\lesssim$ \textbf{10}  & 60 \\ 
	annealed:			 	&500 	& $\lesssim$ \textbf{10}	& 420\\ \hline 
	SMC low energy BM$^*$: 	&$<$ 0.1 	& 28.6 		& 20\\ 
	annealed:		 		&$<$ 1\phantom{0.}  & 17.2 		& 60\\ \hline 
	Crytur: 				&30-50 	& $\lesssim 10$ 		& 60\\ 
	annealed:				&30-50 	& $\lesssim$ \textbf{10} 	&60 \\ \hline 	
	Crytur low energy BM: 	&$<$ 0.1   & 23 		& 15\\ 
	annealed:			 	&$<$ 1\phantom{0.}  & $\lesssim$ \textbf{10}	& 20  \\ \hline 	
	Crytur  high energy BM: &$<$ 2\phantom{0.} & $\lesssim 10$ 		& 10\\ \hline  
	Chemical synthesis:		& $<$ 0.1	& 26.6		& 2 \\ \hline
	
	\end{tabular}
	\caption{Hole widths ($\Gamma$) (bold fonts indicate visible side holes), and lifetimes (T$_a$) of all measured materials at 1.6 K and $B$ = 1 T. $^*$ball milled}
	\label{tab:all}
\end{table}

\begin{figure}[t]
\centering
\includegraphics[width=0.85\columnwidth]{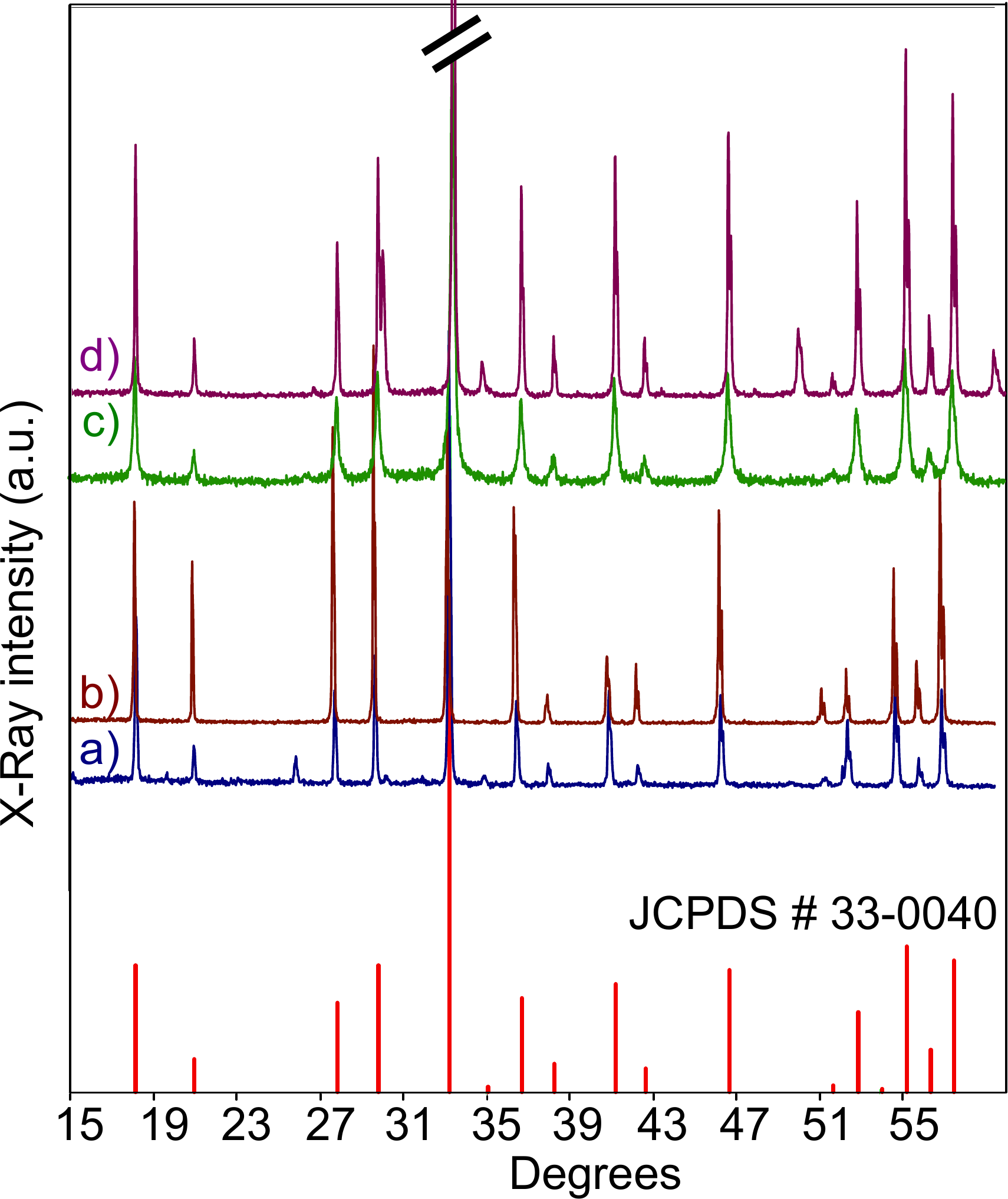}
\caption{XRD spectra of selected powders together with the reference spectrum (JCPDS \# 30-0040) of YAG. a) Chemical synthesis, b) Crytur non-annealed, c) SMC low energy ball milled, d) SMC low energy ball milled and annealed.}
\label{fig:xrd}
\end{figure}

In addition, we performed SEM imaging and powder XRD analysis to further characterize each sample. Example powder XRD spectra are shown in Figure \ref{fig:xrd} for a selected subset of our samples. The spectra were obtained with a Scintag Inc X-1, Advanced Diffraction System. We find that all the analyzed samples formed the expected crystalline structure of YAG. Small deviations from the reference spectrum (JCPDS \# 30-0040) can be observed for some samples. Differences in peak heights can be assigned mostly to sample preparation issues, while additional peaks correspond to the presence of residual amounts of impurity phases \cite{Kareiva2011}.

\subsection{Crushing and ball milling}
\noindent Our first approach to obtain small powders was ``top-down" fabrication starting from a 1\% doped Tm:YAG bulk single crystal originating from the same growth as the one studied above. To pre-crush the crystal, we heated the bulk crystal to $\approx 500$ $^\degree$C and
then immersed the hot crystal in water (21 $^\degree$C) to thermally shock the crystal, causing it to crack into millimeter and larger sized pieces. This procedure was repeated on all pieces with sizes greater than $\approx 5$ mm until all were in the few millimeter or smaller size range. We then ground the small pieces in a mortar and pestle to produce a powder composed of crystallites with sizes of less than 0.5~mm. With spectral hole burning spectroscopy, we measured a hole width of $\lesssim 10$ MHz, limited by power broadening. Side holes due to superhyperfine coupling with $^{27}$Al nuclear spins were still resolvable, as in the bulk material. The spin lifetime was on the order of 1 hour -- significantly shorter than the values obtained in the bulk material at the same temperature and magnetic field. Since this powder originated from a bulk crystal whose properties were well characterized, and was composed of relatively large particles of $\approx 0.5$ mm size that had been ground only for a short time, it is unlikely that any change in chemical composition or in-diffusion of impurities could have occurred (as might be a factor for much smaller particles when processed using high-energy methods).  The only explanation for the significant reduction in lifetime was residual strain induced in the crystallites due to the thermal shocking and grinding used to produce the powder.

\begin{figure}[t]
\centering
\includegraphics[width=\columnwidth]{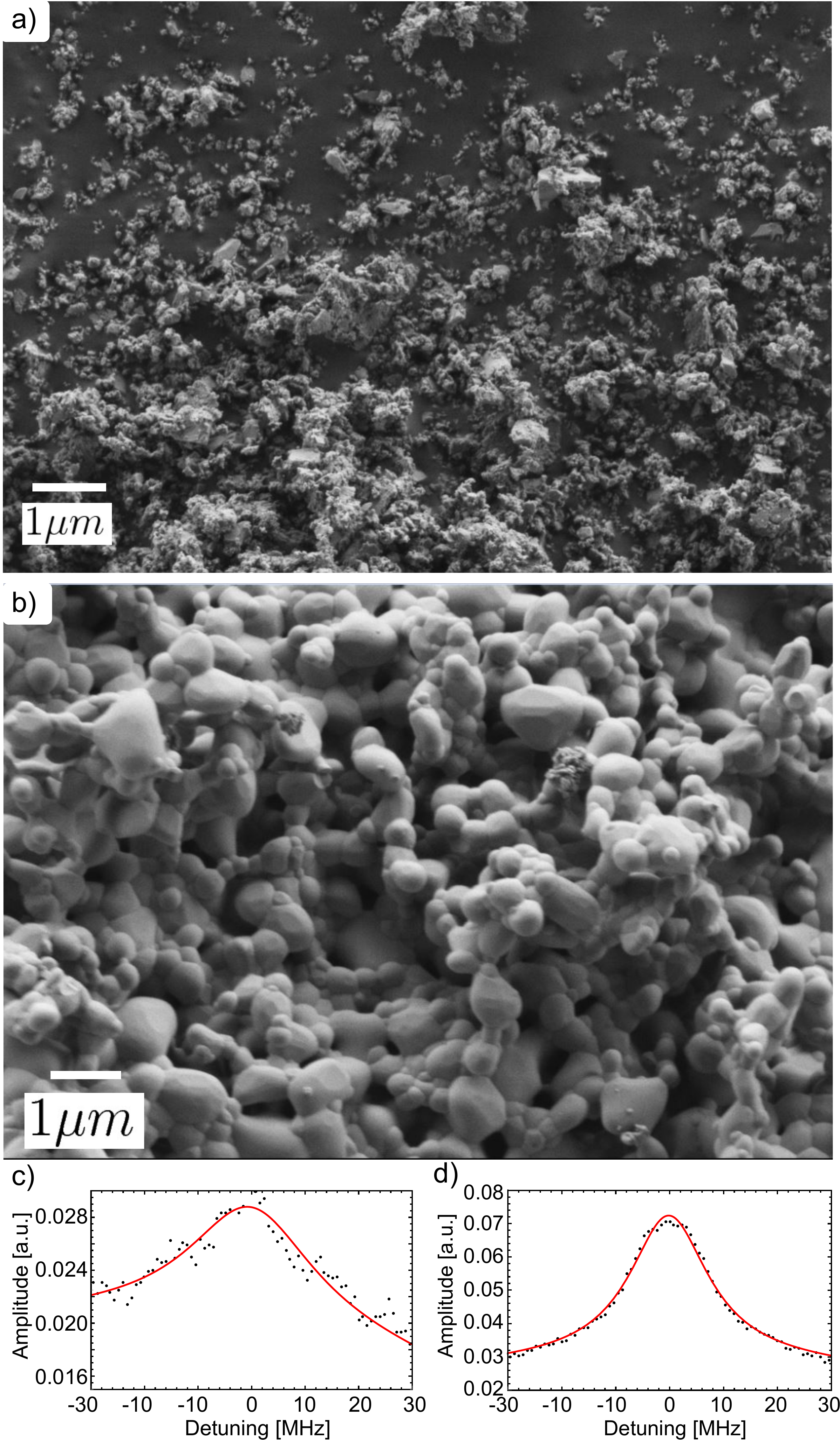}
\caption{SEM images and typical hole burning spectra of the powder obtained after ball-milling the bulk crystal from SMC for two days. (a), (c) before, and (b), (d) after annealing at 1400 $^\degree$C for 4 hours.}
\label{fig:smc-bm}
\end{figure}

With the $< 0.5$ mm sized crushed powder as a starting material, we produced a smaller powder using a low-speed tumbling mill. The Tm:YAG powder was dispersed in ethanol and milled for 48 hours using a mix of zirconia balls with sizes ranging from 1 mm to 1 cm. Figure \ref{fig:smc-bm} (a) shows an SEM image of the obtained powder. The average size of the irregular particles was below 100 nm. The XRD spectrum, shown in Figure \ref{fig:xrd} (c),  displays the expected YAG structure with no observable peak broadening or amorphous background. A spectral hole burning trace, shown in Figure \ref{fig:smc-bm} (c), revealed a hole width of 28.6 MHz. The hole was about 5 times larger than that observed in the bulk, which could not solely be attributed to additional power broadening and reveals a significant difference in the material not detectable from our XRD measurement. 
No sideholes were visible, which was expected since their splitting (10 MHz at $B =$ 1 T) is less than the spectral hole width. Moreover, we found a further reduction of the spin lifetime to 20 minutes. Overall, as opposed to the XRD analysis, the SHB results suggest that a significant amount of damage (strain) was induced in the crystallites during the 2 days of milling.\\

The surprisingly strong effect of mechanical strain on coherence and spin relaxation dynamics may be understood within the framework of the two-level system (TLS) model first developed to describe dynamic disorder modes in amorphous materials \cite{Anderson1972,Phillips1972}. These very low-frequency modes are enabled by the disordered structure and involve groups of atoms tunneling between local configurations with nearly equivalent energies. It has been proposed that low densities of TLS may also be enabled in crystalline
materials by large inhomogeneous lattice strains \cite{Watson1995}, and optical decoherence of rare-earth ions due to interactions with TLS has been observed in both bulk crystals \cite{Flinn1994,Macfarlane2004} and powders \cite{Meltzer1997}. Furthermore, it is known that TLS can also be effective in causing rapid electron and nuclear spin relaxation \cite{Szeftel1978,Askew1986,Balzer-Jollenbeck1988}. Here, both the increase in spectral hole widths and the decrease in nuclear spin lifetimes that we observe in powders are likely to result from the creation of TLS from strain induced during the mechanical fabrication process. This interpretation is also consistent with past observations of increased electronic spin-lattice relaxation rates of Nd$^{3+}$ doped into bulk YAG crystals with greater densities of structural defects in the lattice \cite{Aminov1998}.

\subsection{Effects of annealing}

\noindent Next, we investigated to what degree annealing of the powders can repair the induced damage. After we confirmed that low annealing temperatures around 600 $^\degree$C were unable to improve material properties, and since higher temperatures generally lead to better properties  \cite{Georgescu2007}, we annealed both the 0.5~mm and the 100 nm powders at our highest accessible temperature of 1400 $^\degree$C. The annealing was performed for 4 hours in an oxygen atmosphere (using a tube furnace) to minimize the outdiffusion of oxygen from the YAG matrix. The 0.5 mm annealed crystals appeared unchanged under the SEM. The spin-state lifetime improved significantly to 7 hours and became comparable to that extracted from measurements on the bulk. This suggests that annealing can mostly repair the damages induced by stress during the grinding via mortar and pestle. 

The SEM image of the annealed 100 nm powder, shown in Figure \ref{fig:smc-bm} (b), reveals that the size of the particles increased to around 1 $\mu$m. The particles were almost spherical, with close to uniform size distribution, and appeared to be agglomerated. Compared to the same powder before annealing (panel (a)) this is a major improvement. The XRD spectrum, shown in Figure \ref{fig:xrd} (d), shows no significant change compared to the non-annealed powder. The signal-to-noise ratio in this spectrum is better than the one in the spectrum for the non-annealed powder since more material was available for analysis. This indicates that, in contrary to the SHB measurement, the current XRD analysis is not sensitive enough to detect the changes caused by annealing for this powder. Indeed, the hole burning spectrum in Figure \ref{fig:smc-bm} (d) shows a hole width of 17.2 MHz, which is 30\% smaller than the value measured before annealing. Furthermore, the spin lifetime increased by a factor of three to 1 hour. Hence, our method of annealing also improved the properties of the ball-milled powder, but it was not sufficient to completely re-establish the properties of the bulk. We anticipate that optimized annealing procedures, in particular involving temperatures closer to the melting point, will lead to further improvements.

\subsection{Comparison of high- and low-energy ball-milling methods}
\noindent The above results indicated that the grinding methods used here, namely mortar and pestle and ball-milling, caused damage to the host material. To better understand this surprisingly large effect, and to determine the degree to which damage can be repaired through annealing, we continued studying ball-milling methods starting with a powder from a commercial supplier (Crytur). The powder was created from outer fragments of a large crystal, grown from 99.999\% pure starting materials using the Czochralski method. Those fragments were then pre-crushed using a mechanical jaw crusher and finally milled down to a size of 30-50 $\mu$m in a tumbling mill. The resulting powder was cleaned using mineral acids. 

\begin{figure}[]
\centering
\includegraphics[width=1\columnwidth]{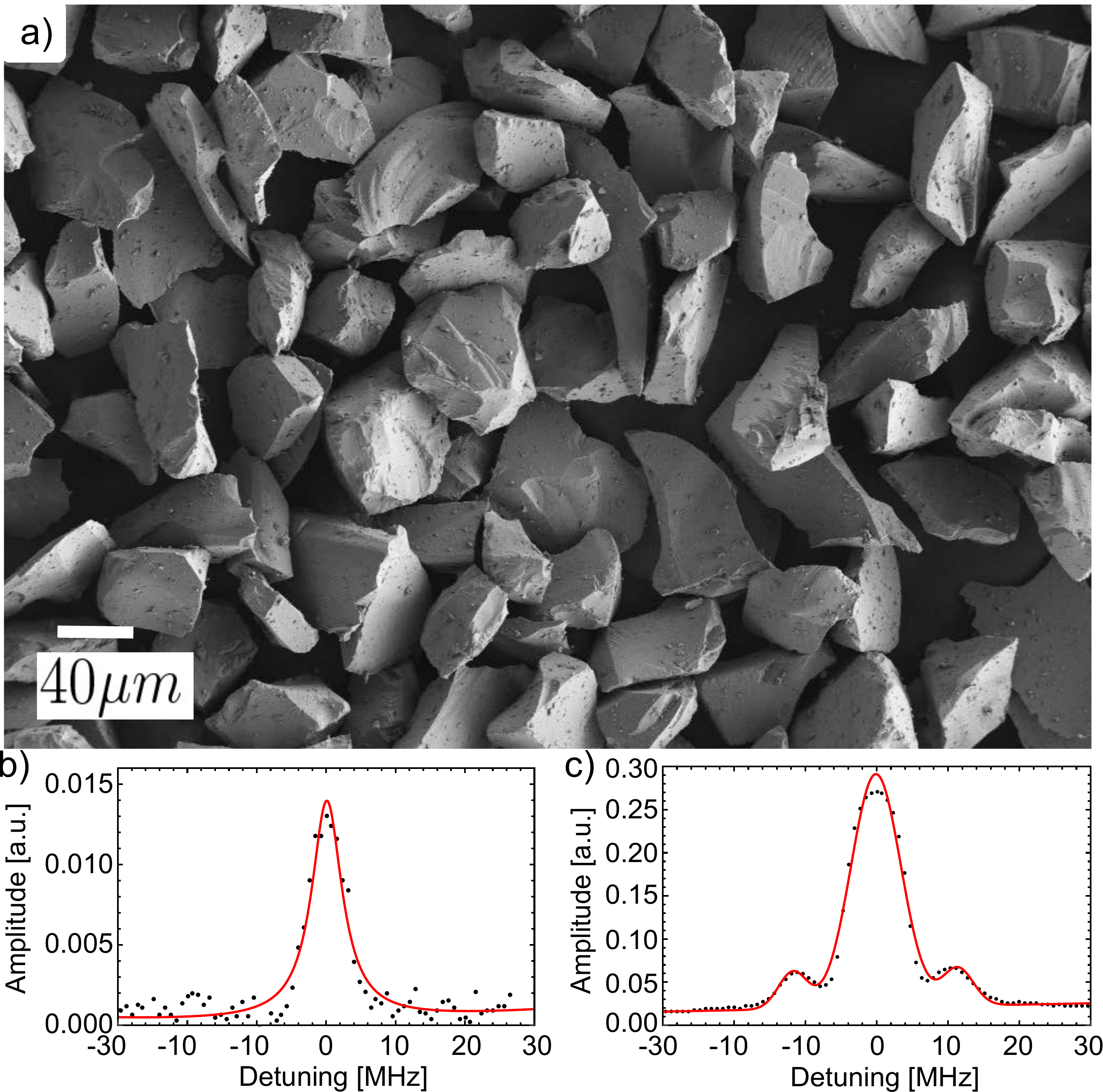}
\caption{(a) SEM image of the 1\% Tm:YAG powder provided by Crytur, and typical hole burning spectrum (b) before and (c) after annealing.}
\label{fig:30um-crytour}
\end{figure}

First, we characterized the ``as purchased" micrometer-sized Crytur powder using  SEM, XRD, and SHB, as shown in Figures \ref{fig:30um-crytour} (a), (b), and \ref{fig:xrd} (b). The SEM image reveals that the powder has sharp edges and few particles with dimensions below 30 $\mu$m. The XRD spectrum shows good YAG crystalline structure. From the hole burning trace we extracted a hole width of $\lesssim$ 10 MHz; however, the lifetimes were only on the order of one hour and comparable to the thermally crushed bulk crystal. This indicates that the crushing method employed by Crytur also induced a comparable amount or strain that is not detected in the XRD spectrum.

Since our annealing procedure was successful for the 0.5 mm-sized powder made from the SMC bulk crystal, we annealed the Crytur powder using the same method. The SEM image taken after annealing resembles that of the powder before annealing (see Figure \ref{fig:30um-crytour} (a)), and the hole width as well as the spin lifetime also remained comparable to those of the powder before annealing. Hence, in contrast to the powder created from the SMC crystal, besides an improvement in signal strength (see Figure \ref{fig:30um-crytour} (c)), our annealing method did not succeed in improving the properties of the powder.
A possible explanation for this observation is the different way of creating the bulk crystal from which the powders were made. While both crystals were grown in a Czochralski process using starting materials of similar purity (99.999\% in the case of Crytur, and 99.995\% for SMC), the original bulk crystal from SMC has been cut from the center of a boule and annealed before shipping. In contrast, the starting material at Crytur was outer fragments of a large crystal that were not annealed before crushing. These differences in the procedures employed in the growth process and sample preparation may affect the density of intrinsic defects and strain in the initial bulk crystal, potentially explaining the differences between the samples that are observed here.

\begin{figure}[t]
\centering
\includegraphics[width=0.9\columnwidth]{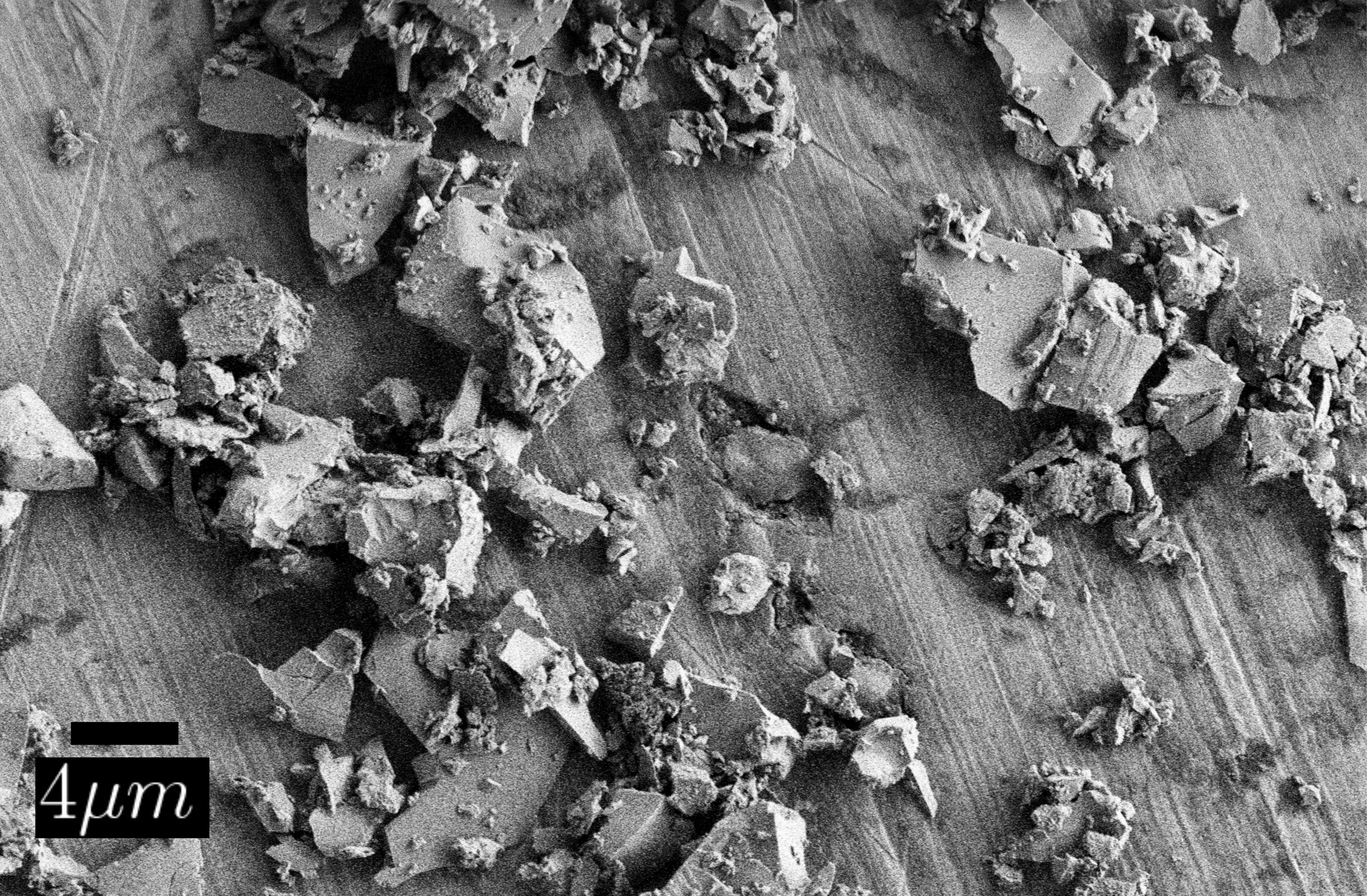}
\caption{SEM image of the 1\% Tm:YAG powder after 4 hours of high-energy planetary ball-milling.}
\label{fig:1um-calgary-hard-ball-mill}
\end{figure}

In addition to jaw crushing and low-energy ball milling, we also used a high-energy planetary ball mill to obtain small powders. This technique requires shorter milling times compared to the low energy approach, but the energy of the impacts during milling is higher. For a direct comparison between high-energy and low-energy ball milling, we used the same starting material, i.e. the non-annealed, 30-50 $\mu$m-size crystals from Crytur, and milled the powder for 4 hours using 1 cm-diameter balls in a high-energy ball-mill. The SEM image in Figure \ref{fig:1um-calgary-hard-ball-mill} showed that the particles were not agglomerated and on average about 2 $\mu$m large. This is the smallest average size we could achieve using this mill with balls of that size. 
Figure \ref{fig:1um-calgary-hard-ball-mill} also reveals the sharp, irregular surfaces of the particles, suggesting significant strain, as well as a broad size distribution. For this powder, we measured a spectral hole width $\lesssim 10$ MHz (no side holes are visible) and a lifetime of around 10 min -- considerably shorter than any other milled material. We conclude that despite the reduced milling time, the crystallites contained an even greater amount of strain and defects due to the high-energy impacts during milling.

Finally, we employed a low-energy tumbling mill for 48 hours to reduce the size of the original Crytur powder. The SEM image reveals agglomerated particles on the order of 100 nm. Furthermore, we found a spectral hole width of 23 MHz (no side holes are visible) and spin lifetimes of around 15 min. This new powder was very similar in all its properties to the one obtained from the SMC bulk crystal after the same ball-milling procedure (see Figures \ref{fig:smc-bm} (a) and (c)). 
The same annealing procedure as described before was then applied to the low-energy ball-milled powder. As observed in the case of the powders originating from the SMC bulk crystal, the particles grew from 100 nm to 1 $\mu$m, exhibiting clear garnet dodecahedral habit. Improvements in terms of spectral hole burning were more pronounced, with a measurement limited hole width of less than 10 MHz and visible side holes. However, the obtained lifetimes were still only around 20 min (as opposed to 1 h in the case of the SMC powder), which might be explained by a larger number of defects or impurities in the original bulk crystal from which the Crytur powder was obtained.

\subsection{Direct chemical synthesis}

\begin{figure}[t]
\centering
\includegraphics[width=0.85\columnwidth]{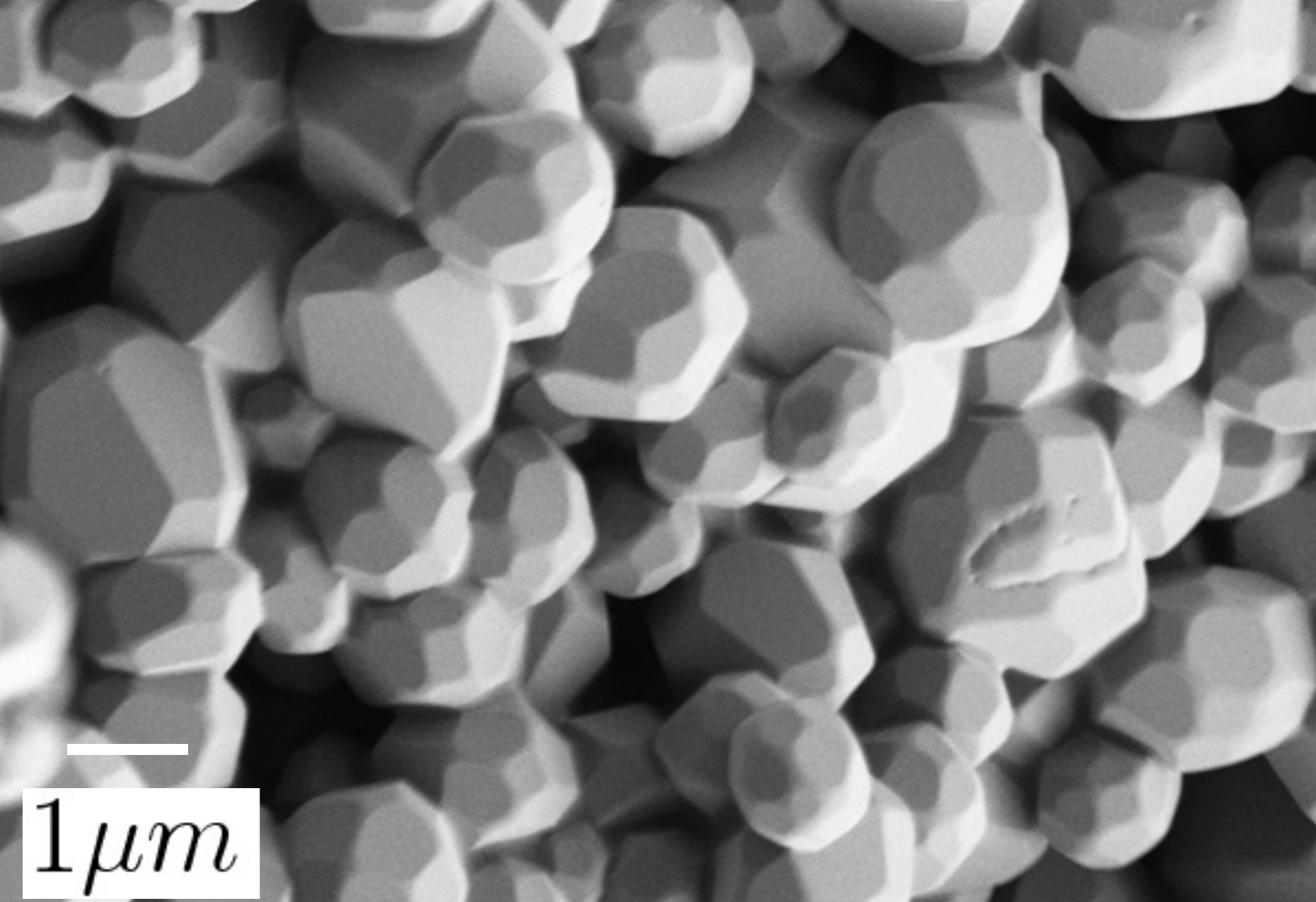} 
\caption{SEM image of the synthesized 1\% Tm:YAG powder. }
\label{fig:1um-calgary-synth}
\end{figure}

\noindent Apart from milling larger crystals, small-sized powders can also be obtained using ``bottom-up" chemical synthesis. In addition to not requiring initial growth of a high-quality crystal -- a substantial advantage for fast prototyping -- this method does not suffer from potential damage induced during grinding. We therefore examined the performance of a 1\% Tm doped YAG powder that was prepared using a co-precipitation method \cite{Li2004,Pradhan2004,Xu2006}. This 1\% Tm:YAG powder was synthesized from starting materials with at least 99.9\% purity. An aqueous nitrate solution was prepared by dissolving Tm(NO$_3$)$_3$~$\cdot$~5H$_2$O, Y(NO$_3$)$_3$~$\cdot$~6H$_2$O, and Al(NO$_3$)$_3$ $\cdot$ 9H$_2$O with a molar ratio of 0.03:2.97:5 in HPLC-grade water. The precipitant solution was prepared with a 1~M concentration of aqueous ammonium bicarbonate and $\approx$1~mM concentrations of sodium dodecyl sulfate and polyethylene glycol 400 added as anionic and nonionic surfactants, respectively. The solutions were mixed by stirring for 1 hour and then the nitrate solution was added drop-wise to an excess of the precipitant solution while stirring at room temperature to form the precipitate. The resulting suspension was aged for several hours while stirring to allow the solutions to fully react, and the resulting gelatinous precipitate was centrifugally separated from the solution. The precipitate was then  washed by vortex mixing and sonication using deionized water, ethanol, and finally n-butanol, and it was then allowed to dry overnight in a vacuum desiccator. A YAG precursor powder was obtained by subsequently drying the precipitate at 300 $^\degree$C for 4 hours and then grinding in a mortar to obtain a fine powder. The precursor was then mixed with 5\% Li$_2$CO$_3$ by weight as a flux to promote growth of large crystallites, placed in a covered alumina crucible, and then crystallized in a tube furnace at a temperature of 1300 $^\degree$C for 8 hours in air. Analysis of the resulting powder using XRD, shown in Figure \ref{fig:xrd} (a), revealed complete formation of single-phase crystalline YAG. The quality of the XRD spectrum, in terms of signal-to-noise ratio and peak widths, is comparable to the spectrum of the purchased Crytur powder (see Figure~\ref{fig:xrd} (b)). 
An SEM image of the synthesized 1\% Tm:YAG powder is shown in Figure \ref{fig:1um-calgary-synth}, revealing that it consisted of uniform, non-agglomerated crystallites with diameters of 500 nm to 2 $\mu$m that exhibited the characteristic dodecahedral crystal habit of the garnets.

Although the XRD and SEM results for the synthesized powders appeared promising, spectral hole burning measurements revealed the poor quality of the material. We observed a very broad hole width of 26.6 MHz with no visible side holes. Furthermore, the measured nuclear spin lifetime was limited to values of at most 2 minutes. Hence, the properties of the synthesized powder were significantly worse than the properties of all of the powders obtained using the top-down ball-milling approach. We expect that the poor performance of the synthesized crystallites may result from residual chemical impurities that can cause lattice strain from incorporation of the defects as well as the presence of paramagnetic impurities such as iron that can couple to the Tm$^{3+}$ ions and induce nuclear spin relaxation and optical decoherence. Consequently, further studies are required on powders synthesized from higher-purity starting materials to investigate whether the properties show improvement and how they compare to powders of similar purity obtained from top-down fabrication.

\section{Discussion of different characterization methods}

The three methods used in this work, SEM, XRD and SHB, complement each other as they provide information about different aspects of the nanocrystal properties. We found that SHB was the most sensitive method to detect effects of crushing as well as annealing on the powder quality in our measurements.

SEM images constitute the most convenient method to assess the size of the particles, and also observe their shape, which can be an indication of the crystalline quality, provided one knows the crystal's habit. For instance, the powder obtained by chemical synthesis exhibits the characteristic dodecahedral crystal habit of YAG (see Figure~\ref{fig:1um-calgary-synth}). However, the SEM results do not allow one to identify the chemical composition, and thus cannot detect impurities present in the particles.

XRD analysis is the most reliable method to assess the composition and the phase of the nanocrystals. If the concentration of impurities is high enough, they can also be detected by this method. Potential amorphous character of the particles should also be visible as a broad background on the XRD spectrum. However, we were not able to observe the amorphous components or crystal strain that was clearly observed in our SHB measurements. In principle, by analyzing the peak widths in the spectrum, XRD can also give access to the particle size, but this requires a resolution beyond that of our instrument.

The SHB measurement gives access to the population lifetime of the nuclear spin states of Tm$^{3+}$:YAG, as well as to the homogeneous linewidth of the optical transition if the hole width is not broadened by laser power, which limited our resolution to 10 MHz. Both quantities are very sensitive to the crystal quality in terms of both strain and impurities. This sensitivity originates from the coupling between the ion's nuclear spin and the environment. In some cases, the SHB measurements reveal differences in the materials that are not observed in the SEM or XRD results. For example, the chemically synthesized nanocrystals exhibited very short spin lifetimes despite promising SEM and XRD results. The SHB measurement gives a quantitative, and sensitive method to determine the overall quality of the nanocrystals, even though it does not directly determine the nature of the limiting factors.

\section{Conclusions and Outlook}

\noindent In summary, we studied properties of 1\% Tm:YAG powders produced by different methods via SEM imaging, XRD analysis, and spectral hole burning. We found that SHB is a very sensitive and well suited method to assess the quality of REI doped crystals in a quantitative way.

We found that any grinding or milling technique degraded the performance of the REI doped material in terms of homogeneous linewidth and especially spin-state lifetimes, most likely due to induced stress and deformation to the host matrix. Annealing is crucial and allowed us to recover the bulk spin-state lifetimes in the case of a soft-grinding method. In the case of more severe damage caused by ball-milling, bulk properties were partly recovered in powders originating from high-quality, annealed bulk crystals, whereas it remained poor in the case of powders obtained from non-annealed crystal fragments. This shows the importance of starting with high-quality crystals and the need for annealing. Finally, we found that the performance of the synthesized powder studied here was likely limited by the quality of the starting materials. 

We anticipate our findings to be generalizable to other impurities and other hosts, in particular to color centers in nano-structured diamond, and to provide valuable insight towards nano-structuring REI doped crystals. Furthermore, SHB is a way to reveal variations in terms of quality and uniformity of powders that could have consequences for traditional applications where low levels of lattice defects can affect performance, such as luminescence efficiency and thermal quenching of phosphors.\\

\noindent The authors acknowledge support from Alberta Innovates Technology Futures (ATIF), the National Engineering and Research Council of Canada (NSERC), and the National Science Foundation of the USA (NSF) under award nos. PHY-1212462, PHY-1415628, and CHE-1416454. W. T. is a Senior Fellow of the Canadian Institute for Advanced Research (CIFAR). This work is based upon research conducted in part at the Montana State University Imaging and Chemical Analysis Laboratory.


\begin{thebibliography}{10}
\newcommand{\enquote}[1]{``#1''}

\bibitem{Riedmatten2015}
H.~de~Riedmatten and M.~Afzelius, \enquote{Quantum light storage in solid state
  atomic ensembles},   ArXiv:1502.00307 (2015).

\bibitem{Saglamyurek2014}
E.~Saglamyurek, N.~Sinclair, J.~A. Slater, K.~Heshami, D.~Oblak, and W.~Tittel,
  New Journal of Physics \textbf{16}, 065019 (2014).

\bibitem{Menager2001}
L.~M\'{e}nager, I.~Lorger\'{e}, J.-L.~L. Gou\"{e}t, D.~Dolfi, and J.-P.
  Huignard, Optics Letters \textbf{26}, 1245 (2001).

\bibitem{mcauslan_strong-coupling_2009}
D.~L. McAuslan, J.~J. Longdell, and M.~J. Sellars, Physical Review A
  \textbf{80}, 062307 (2009).

\bibitem{obrien_interfacing_2014}
C.~O\textquotesingle Brien, N.~Lauk, S.~Blum, G.~Morigi, and M.~Fleischhauer, Physical Review
  Letters \textbf{113}, 063603 (2014).

\bibitem{walther_high_2015}
A.~Walther, L.~Rippe, Y.~Yan, J.~Karlsson, D.~Serrano, A.~N. Nilsson,
  S.~Bengtsson, and S.~Kr\"oll, ArXiv:1503.08447 (2015).

\bibitem{lutz_modification_2015}
T.~Lutz, L.~Veissier, C.~W. Thiel, R.~L. Cone, P.~E. Barclay, and W.~Tittel ,ArXiv: 1504.02471 (2015).
  .

\bibitem{Hong1998}
K.~Hong, R.~Meltzer, B.~Bihari, D.~Williams, and B.~Tissue, Journal of
  Luminescence \textbf{76–77}, 234  (1998). Proceedings of the Eleventh
  International Conference on Dynamical Processes in Excited States of Solids.

\bibitem{malyukin_single-ion_2003}
Y.~V. Malyukin, A.~A. Masalov, and P.~N. Zhmurin, Physics Letters A
  \textbf{316}, 147 (2003).

\bibitem{Macfarlane2001}
R.~M. Macfarlane and M.~J. Dejneka, Optics Letters \textbf{26}, 429 (2001).

\bibitem{perrot_narrow_2013}
A.~Perrot, P.~Goldner, D.~Giaume, M.~Lovri\'c, C.~Andriamiadamanana, R.~R.
  Gon\c calves, and A.~Ferrier, Physical Review Letters \textbf{111}, 203601
  (2013).

\bibitem{Heitjans2007}
P.~Heitjans, M.~Masoud, A.~Feldhoff, and M.~Wilkening, Faraday Discussions
  \textbf{134}, 67 (2007).

\bibitem{Scholz2002}
G.~Scholz, R.~St\"osser, J.~Klein, G.~Silly, J.~Y. Buzaré, Y.~Laligant, and
  B.~Ziemer, Journal of Physics: Condensed Matter \textbf{14}, 2101 (2002).

\bibitem{sun_symmetry_2000}
Y.~Sun, G.~M. Wang, R.~L. Cone, R.~W. Equall, and M.~J.~M. Leask, Physical
  Review B \textbf{62}, 15443 (2000).

\bibitem{macfarlane_coherent_1987}
R.~M. Macfarlane and R.~M. Shelby, in \emph{Modern {Problems} in {Condensed}
  {Matter} {Sciences}}, vol.~21 of \emph{Spectroscopy of {Solids} {Containing}
  {Rare} {Earth} {Ions}}, A.~A. Kaplyanskii and R.~M. Macfarlane, eds.
  (Elsevier, 1987), pp. 51--184.

\bibitem{bulk-paper}
L.~Veissier, T. Lutz, C.~W. Thiel, R.~L. Cone, and W.~Tittel, in preparation.

\bibitem{Liu2005}
G.~Liu and B.~Jacquier, \emph{Spectroscopic Properties of Rare Earths in
  Optical Materials} (Springer Berlin Heidelberg, 2005).

\bibitem{Bonarota2010}
M.~Bonarota, J.~Ruggiero, J.~L.~L. Gou\"et, and T.~Chaneli\`ere, Physical Review A
  \textbf{81}, 033803 (2010).
  
\bibitem{Kareiva2011}
A.~Kareiva, Materials Science 17, 1392-1320 (2011).

\bibitem{Anderson1972}
P.~W. Anderson, B.~I. Halperin, and C.~M.~Varma, Philosophical Magazine
  \textbf{25}, 1 (1972).

\bibitem{Phillips1972}
W.~Phillips, Journal of Low Temperature Physics \textbf{7}, 351 (1972).

\bibitem{Watson1995}
S.~K. Watson, Physical Review Letters \textbf{75}, 1965 (1995).

\bibitem{Flinn1994}
G.~P. Flinn, K.~W. Jang, J.~Ganem, M.~L. Jones, R.~S. Meltzer, and R.~M.
  Macfarlane, Physical Review B \textbf{49}, 5821 (1994).

\bibitem{Macfarlane2004}
R.~M. Macfarlane, Y.~Sun, R.~L. Cone, C.~W. Thiel, and R.~W. Equall, Journal of
  Luminescence \textbf{107}, 310  (2004). Proceedings of the 8th International
  Meeting on Hole Burning, Single Molecule, and Related Spectroscopies: Science
  and Applications.

\bibitem{Meltzer1997}
R.~Meltzer, K.~Jang, K.~Hong, Y.~Sun, and S.~Feofilov, Journal of Alloys and
  Compounds \textbf{250}, 279  (1997).

\bibitem{Szeftel1978}
J.~Szeftel and H.~Alloul, Journal of Non-Crystalline Solids \textbf{29}, 253
  (1978).

\bibitem{Askew1986}
T.~R. Askew, H.~J. Stapleton, and K.~L. Brower, Physical Review B \textbf{33}, 4455
  (1986).

\bibitem{Balzer-Jollenbeck1988}
G.~Balzer-J\"ollenbeck, O.~Kanert, J.~Steinert, and H.~Jain, Solid State
  Communications \textbf{65}, 303  (1988).

\bibitem{Aminov1998}
L.~Aminov, I.~Kurkin, and D.~Lukoyanov, Applied Magnetic Resonance \textbf{14},
  447 (1998).

\bibitem{Georgescu2007}
S.~Georgescu, S.~Constantinescu, A.~Chinie, A.~Stefan, O.~Toma, and I.~Bibicu,
  in \emph{ROMOPTO 2006: Eighth Conference on Optics} (International Society
  for Optics and Photonics, 2007), pp. 67850A--67850A.
  
\bibitem{Li2004}
X.~Li, H.~Liu, J.~Wang, H.~Cui, X.~Zhang, and F.~Han, Materials Science and
  Engineering: A \textbf{379}, 347  (2004).

\bibitem{Pradhan2004}
A.~Pradhan, K.~Zhang, and G.~Loutts, Materials Research Bulletin \textbf{39},
  1291  (2004).

\bibitem{Xu2006}
G.~Xu, X.~Zhang, W.~He, H.~Liu, H.~Li, and R.~I. Boughton, Materials Letters
  \textbf{60}, 962  (2006).

\end{thebibliography}
\end{document}